\documentstyle[preprint,aps]{revtex} 

\newcommand{\be}{\begin{equation}}
\newcommand{\ee}{\end{equation}}

\newcommand{\bea}{\begin{eqnarray}}
\newcommand{\eea}{\end{eqnarray}}

\newcommand{\ba}{\begin{array}}
\newcommand{\ea}{\end{array}}

\begin{document}

\pagenumbering{arabic} 

\title{ Berry Phase and Dissipation of Topological Singularities }

\author{  Ping Ao and  Xiao-Mei Zhu \\
       Ume\aa{\ }University, Ume\aa, Sweden  }

\maketitle

\begin{abstract} 
We outline a path integral derivation of both the transverse force, 
the Berry phase, and friction for a vortex from the microscopic fermionic
superfluid theory.
The derivation manifests transparently the mutual
independence of the Berry phase and dissipative terms 
in the effective vortex action. \footnote{talk given at 
Solitons: Properties, Dynamics, Interactions, Applications, 
July 20-26, 1997, Queen's University, Kingston, Canada }
\end{abstract}  

\section{ Introduction } 
\label{starting}  

Vortices in superconductors and He3 superfluids are topological 
excitations. They determine the global properties such as
the stability of the supercurrent carrying states, 
and have been under intensive theoretical and experimental 
studies since earlier sixties\cite{brandt}.
The detailed microscopic Bardeen-Cooper-Schrieffer type (BCS) theory
for fermionic superfluids is successful and well defined. 
The derivation of vortex dynamics is, however,
non-trivial and less certain.  
Here we present a path integral derivation of vortex 
dynamics following the line of
geometric methods\cite{ao1,thouless}.
The advantage of the present method is that in the effective vortex Lagrangian
the separation of geometric and dissipative contributions 
is natural. The important results here are  
that the transverse force agrees with the one obtained by 
the Berry phase method\cite{ao1} and by the total
force-force correlation function method\cite{thouless}, 
and is insensitive to details. 
The friction is determined by the
spectral function of the Hamiltonian, sensitive to details. 
The large transverse force has been found recently by 
a direct measurement\cite{zhu}, in consistent with the prediction in
Ref.\cite{ao1,thouless}.

\section{ Description of the Berry Phase and Dissipation }

In this section we review the description of the
Berry phase and dissipation 
for a particle within a model Hamiltonian, and show how those two 
quantities appear independently in the effective particle action. 

To study the quantum dissipative dynamics of a particle, 
one may specify the whole Hamiltonian, particle plus environment.
In practice, what normally needed is the particle dynamics with the influence
from the environment. The environment itself is left unobserved.
Hence an accurate and efficient method to 
integrate out environmental degrees of freedoms
is needed. This can be most readily done using the path integral method. 
In the present paper
the approach and notation of Ref.\cite{leggett} to this question will
be followed. In that formulation, the dissipation is produced
by an environment consisting of a set of harmonic oscillators.
The corresponding total model Hamiltonian may have the following form:
\be
   H =   \frac{1}{2m_v} \left[ {\bf P} + {\bf A}({\bf x }) \right]^{2}
       + U({\bf x })  
   + \sum_{j} 
       \left[ \frac{1}{2m_{j} } {\bf p}_{j}^{2} + \frac{1}{2} m_{j}
       \omega_{j}^{2} \left(  {\bf q}_{j} - \frac{c_{j} }{m_{j}\omega_{j}^{2} }
                        {\bf x } \right)^{2} \right] \; .
\ee
Here ${\bf x} = (x,y), {\bf P},  \{{\bf q}_j\}$,  and $\{{\bf p}_j \} $ 
are all two dimensional vectors. 
The vector potential ${\bf A}$ determined by 
$ \nabla\times {\bf A} = {\bf B} $ corresponds  the transverse force
$ - \dot{\bf x}\times {\bf B}$, with {\bf B} in the z-direction.
The effect of the dissipative 
environment is specified by the spectral function
\be
   J(\omega) \equiv \frac{\pi}{2} \sum_{j} \frac{c_{j}^{2} }{m_{j}\omega_{j} } 
               \delta(\omega - \omega_{j} ) \; .
\ee
In the present paper, we shall assume the spectral function to have the 
following form
\be
   J(\omega) = \eta \omega^{s} 
         \exp\left\{ - \frac{\omega}{\omega_{c} } \right\} \; ,
\ee 
with $\omega_{c}$ the cutoff frequency whenever needed.
In accordance with Ref.\cite{leggett},
$s> 1 $ is the superohmic case, $s=1$ the ohmic case, and $0\leq s<1$ 
the subohmic case. 
In the ohmic damping case, $\eta$ is the friction coefficient in the usual 
Langevin equation.

In the imaginary time formulation 
the particle  dynamics is described by the Euclidean action
\bea
   S & = & \left. \int_{0}^{\hbar\beta } d\tau \right[ 
     \frac{1}{2} m_v \dot{\bf x}^{2} + 
      i \dot{\bf x } \cdot {\bf A } +  U({\bf x}) \nonumber \\
   & & +  \sum_{j} 
     \left.  \left( \frac{1}{2} m_{j} \dot{\bf q}_{j}^{2} 
              + \frac{1}{2}m_{j}\omega_{j}^{2} \left( {\bf q}_{j} 
                           - \frac{c_{j} }{m_{j} \omega_{j}^{2} } 
                       {\bf x} \right)^{2} \right) \right] \; ,
\eea
where $\beta = 1/k_{B}T$ is the inverse temperature.

Since we are interested in the particle dynamics, 
keeping other degrees of freedoms of the environment is unnecessary. 
The integrations over $\{ q_{j} \}$ will then be taken.
The resulting effective action for the particle is\cite{leggett,at}
\bea
      S_{eff}[ {\bf x}(\tau) ] & = & \int_{0}^{\hbar\beta } d\tau 
    \left[ \frac{1}{2} m_v \dot{\bf x}^{2} + 
      i \dot{\bf x } \cdot {\bf A} + U({\bf x}) \right]  \nonumber \\
  & & + \frac{1}{2} \int_{0}^{\hbar\beta } d\tau \int_{0}^{\tau } d\tau'
      F_{\parallel}(\tau - \tau' ) [ {\bf x}(\tau) - {\bf x}(\tau') ]^{2} \; ,
\eea
with the damping kernel $F_{\parallel}$ as
\be 
   F_{\parallel}(\tau) = \frac{1}{\pi} \int_{0}^{\infty} d\omega \;  J(\omega) 
   \frac{ \cosh\left[\omega\left(\frac{\hbar\beta}{2} - |\tau|\right)\right] }
             { \sinh\left[\frac{\omega\hbar\beta}{2} \right] } \; .
\ee
It is evident that the transverse force corresponds the term 
$i \dot{\bf x } \cdot {\bf A}$ in the (effective) particle action.
It has not been influenced 
by the dissipative environment, due to the rotational and translational 
invariances of the coupling between the particle and the environment.
This term gives the Berry phase, 
or the Aharonov-Bohm phase for a charged particle,
if moving along a closed trajectory.
The influence of the environment is contained in the non-local term. 
It should be pointed out that in the dissipative dynamics described by 
Eqs.(5,6) all one needs to know is the spectral function $J(\omega)$. 
The details of the original environmental Hamiltonian
have been suppressed. This implies that 
the dissipation can be produced by a fermionic environment, 
an observation important in our following derivation.

\section{ Effective Vortex Action }

To find the effective vortex action,
we begin with the standard BCS Lagrangian for s-wave pairing
in the imaginary time representation. Here 
the unwanted fermionic degrees of freedoms will be integrated out,
instead of those bosonic ones in section 2.
We will only consider neutral superfluids here,
but the coupling to electromagnetic fields does no affect our main results. 
More detailed analysis will be published elsewhere.\cite{az}
The model BCS  Lagrangian is
\begin{eqnarray}
   L_{BCS} & = &  \sum_{\sigma}\psi^{\dag}_{\sigma}( x,\tau) 
      \left(  \hbar\partial_{\tau}- \mu_F 
    - \frac{\hbar^{2}}{2m} \nabla^{2} + V(x)  \right) \psi_{\sigma}(x,\tau)
       \nonumber \\
      & & - g\psi^{\dag}_{\uparrow}(x,\tau) \psi^{\dag}_{\downarrow}(x,\tau)
          \psi_{\downarrow}(x,\tau) \psi_{\uparrow}(x,\tau )  \; , 
\end{eqnarray}
where $\psi_{\sigma}$ describes electrons with spin $\sigma=(
\uparrow, \downarrow )$, $\mu_F$ the chemical potential determined by the 
electron number density, $V(x) $ the impurity potential, and $x =(x,y,z)$.
A vortex at $x_v$ has been implicitly assumed. 
The partition function is 
\be
   Z =  \int {\cal D}\{x_v, \psi^{\dag}, \psi \} \times             
           \exp\left\{ - \frac{1}{\hbar} \int_0^{\hbar\beta } 
             d\tau \int d^3x L_{BCS} \right\}   \; ,      
\ee
with $\beta = 1/k_B T $ , and $d^3x = dxdydz$.
Inserting the identity in the functional space,
\[ 
   1 =  \int {\cal D}\{ \Delta^{\ast}, \Delta \}
   \exp \left\{- \frac{g}{\hbar} \int_0^{\hbar\beta } d\tau \int d^3x\times
       \left| \psi_{\downarrow} \psi_{\uparrow}
     + \frac{1}{g} \Delta (x,\tau) \right|^2  \right\} \; ,
\]
into Eq.(8) we have 
\bea
   Z  &  =  & \int {\cal D}\{x_v, \psi^{\dag}, \psi, 
        \Delta^{\ast}, \Delta \} \times     
      \exp \left\{ 
        - \frac{1}{\hbar} \int_0^{\hbar\beta } d\tau \int d^3x
           \right. \times  \nonumber \\
    &  &  \left( \psi^{\dag}_{\uparrow}, 
              \psi_{\downarrow} \right )
       ( \hbar\partial \tau + {\cal H} )  \left( \begin{array}{c} 
                       \psi_{\uparrow}   \\          
                       \psi^{\dag}_{\downarrow}
                       \end{array}  \right)  
      \left. - \frac{1}{g}  \int_0^{\hbar\beta } d\tau \int d^3x  
         |\Delta |^2   \right\} \; . \nonumber
\eea
Here the Hamiltonian is defined as
\be
   {\cal H}( \Delta, \Delta^{\ast} ) = \left( \begin{array}{cc} 
                     H & \Delta  \\
                    \Delta^{\ast} & - H^{\ast} 
                   \end{array} \right) \; ,
\ee
with $H =  - (\hbar^{2}/2m ) \nabla^{2} -  \mu_F + V(x)$.
Integrating out the electron
fields $\psi_{\sigma}^{\dag}$ and $ \psi_{\sigma}$ first, then 
the auxiliary(pair) fields $\Delta$ under the meanfield approximation, 
one obtains the partition function for the vortex 
\be   
   Z = \int {\cal D}\{ x_v \}
       \exp \left\{ - \frac{ S_{eff} }{ \hbar }  \right\} \; ,
\ee 
with the effective vortex action
\be
   \frac{S_{eff} }{\hbar} = - Tr \ln G^{-1} -
     \frac{1}{\hbar g}\int_0^{\hbar\beta} d\tau \int d^{3}x|\Delta|^{2} \; ,
\ee
where $Tr$ includes internal and  space-time indices,
and the Nambu-Gor'kov (NG) Green's function $G$ defined by 
\be
     ( \hbar \partial_\tau + {\cal H})
      G(x,\tau; x',\tau') = \delta(\tau-\tau') \delta^3(x-x') , 
\ee
together with the BCS gap equation, or the self-consistent equation,
\be
     \Delta(x,\tau) = -  g \;   
   < \psi_{\downarrow}(x,\tau) \psi_{\uparrow}(x,\tau) > \; .
\ee
A special attention should be paid to the equal time limit of the 
NG Green's function.\cite{schrieffer}

We assume that the vortex is confined to move in a small regime around a point 
at $x_0$, 
which allows a small parameter expansion in terms of the difference 
between the vortex position $x_v$ and $x_0$. 
We look for the long time behavior of vortex dynamics under this small
parameter expansion. 
For the meanfield value of the order parameter, this expansion is
\be
   \Delta(x,\tau, x_v) =  \left( 1 
         + \delta x_v(\tau) \cdot \nabla_0 
       + \frac{1}{2}  ( \delta x_v(\tau) \cdot \nabla_0 )^2 \right) 
         \Delta_0(x,x_0) \; .
\ee
Here $\delta x_v = x_v - x_0$.
In Eq.(14) we have used the fact that when $x_v = x_0$ 
the vortex is static.
The effective vortex action to the same order is, 
after dropping a constant term,
\be
     \frac{S_{ eff } }{\hbar }  =  - \frac{1}{2} Tr (G_0 \Sigma' )^2 
           + \frac{1}{\hbar g} \int_0^{\hbar\beta} d\tau\int d^3x \times 
         \delta x_v\cdot \nabla_0 \Delta^{\ast}_0 \; 
                 \delta x_v\cdot \nabla_0 \Delta_0 \; ,
\ee
with
\be
    \Sigma' = \left( \begin{array}{cc}
                    0 &  \delta x_v\cdot \nabla_0 \Delta_0 \\
             \delta x_v\cdot \nabla_0 \Delta^{\ast}_0 & 0 
                    \end{array} \right)  \; .
\ee
Here $G_0$ is the NG Green's function with $\Delta(\Delta^{\ast})$
replaced by  $\Delta_0(\Delta^{\ast}_0)$,
and the gradient $\nabla_0$ is with respect to $x_0$.

Before constructing the NG Green's function $G_0$  
we consider the eigenfunctions of the Hamiltonian 
${\cal H}_0 = {\cal H}(\Delta_0, \Delta^{*}_0) $. 
The stationary 
Schr\"{o}dinger equation, the Bogoliubov-de Gennes (BdG) equation, is 
\be
   {\cal H}_0 \Psi_k(x) = E_k \Psi_k(x) \; ,
\ee
with 
\[
   \Psi_k(x) = \left( \begin{array}{c} u_k(x) \\ v_k(x) \end{array} 
       \right) \; .
\]   
Since ${\cal H}_0$ is hermitian, all its eigenstates form a  
complete and orthonormal set. Eq.(17) have two useful properties: 
\be
    {\cal H}_0 \Psi(x) = E \; \Psi(x)     \Rightarrow 
    {\cal H}_0 \overline{\Psi}(x) = - E \; \overline{\Psi}(x) \; ,
\ee
with 
\[
  \overline{\Psi}(x) = \left( \begin{array}{c} 
                                v^{\ast}(x) \\ 
                      - u^{\ast}(x)\end{array} \right) \; ,
\]
and, since $\delta x_v \cdot \nabla_0 {\cal H}_0 = \Sigma'$, 
for $k\neq k'$, 
\be
    \int d^3x \Psi_k^{\dag}(x)  \Sigma' \Psi_{k'}(x) \\
      =  (E_{k'} - E_{k} ) \delta x_v \cdot \int d^3x \Psi_k^{\dag}(x)  
      \nabla_0\Psi_{k'}(x)  \; .
\ee
This equation implies that the two ways of specifying the vortex 
coordinate, through the trapping potential or through the order parameter,
are equivalent. 

The NG Green's 
$G_0$ can be expressed as
\be
   G_0(x,\tau; x',\tau' ) = \sum_{n,k} \frac{-1}{\hbar \beta} 
          \frac{ e^{ - i \omega_n (\tau -\tau') } }{ i\hbar \omega_n - E_k }
          \Psi_k(x) \Psi_k^{\dag}(x') .
\ee
Here $\omega_n = n \pi / \hbar\beta$, with $n$ odd integers.
Assuming the rotational symmetry after the impurity average,
a straightforward calculation leads to the following effective vortex action 
\bea
    S_{eff} & = & \left. \frac{1}{2}\int_0^{\hbar\beta} d\tau 
     \right[ K \; |\delta x_v(\tau) |^2 
               +  \int_0^{\tau } d\tau' 
      F_{\parallel} ( \tau-\tau')|\delta x_v(\tau)- \delta x_v(\tau') |^2 
       \nonumber \\
           & &   + \left.  \int_0^{\hbar\beta} d\tau' 
            F_{\perp} ( \tau-\tau') 
           (\delta x_v(\tau) \times \delta x_v(\tau') )\cdot \hat{z} 
                  \right]  \; ,
\eea
with the spring constant in the effective potential,
\be
    K =  \frac{1}{g} \int d^3x |\nabla_0 \Delta_0^{\ast}(x,x_0)|^2
        - \int^{\infty}_{0} d\omega \frac{ J(\omega) }{\omega} \; ,
\ee
the damping kernel,
\be
    F_{\parallel}(\tau) = \frac{1}{\pi} \int^{\infty}_{0} d\omega J(\omega)
    \frac{\cosh\left[\omega\left(\frac{\hbar\beta }{2}-|\tau|\right)\right] }  
         {\sinh\left[\omega \frac{\hbar\beta }{2} \right]          }  \; ,
\ee
and the transverse kernel, in the long time limit,  
in terms of the virtual transitions,
\bea
   F_{\perp}(\tau) & = & - \partial_{\tau-\tau'} \delta(\tau-\tau') 
                       \sum_{k,k'} \int d^3x \int d^3x' \;
        \hbar ( f_k - f_{k'} ) \nonumber \\
    & &   \frac{1}{2} \hat{z}\cdot
       \left(  \Psi_k^{\dag}(x') \nabla_0\Psi_{k'}(x')
        \times \nabla_0\Psi_{k'}^{\dag}(x)  \Psi_k(x) \right) \; ,
\eea
or in terms of the contribution from each state,
\bea
   F_{\perp}(\tau) & = & - \partial_{\tau-\tau'} \delta(\tau-\tau') \sum_{k}
                                    \int d^3x \; \hbar \hat{z}\cdot 
  \left( f_k \nabla_0u_k^{\ast}(x)\times \nabla_0 u_k(x) \right. \nonumber \\
    & &   
   \left.  - (1-f_k) \nabla_0v_k^{\ast}(x)\times \nabla_0 v_k(x) \right)  \; .
\eea
To reach Eqs.(21-25), following two identities have also been used:
\be
     \sum_n \frac{ e^{ -i \omega_n \delta}  }
                 { i\hbar \omega_n - E_k    } = 
     \left\{
        \ba{cc}
       \beta \; f_k \; ,      & \delta = 0^-  \\
       - \beta \; (1 - f_k )\; , & \delta = 0^+  \\    
         \ea
   \right. \; ,
\ee
and
\be
         \sum_{n-n'} \frac{ - 1}{\beta} 
      \frac{ e^{ - i (\omega_n - \omega_{n'}) \tau} }
           { i \hbar(\omega_n - \omega_{n'} ) - E   } 
        = \frac{1}{2} 
  \frac{\cosh\left[\frac{E }{\hbar}\left(\frac{\hbar\beta}{2}
                                                - |\tau|\right)\right] }
        {\sinh\left[ E \frac{\beta}{2} \right]             }  \;,
\ee    
with the Fermi distribution function $f_k = 1/(1 + e^{\beta E_k} )$, and the 
spectral function
\bea
   J(\omega) & = & \frac{\pi }{2} \sum_{k,k'} 
     \delta(\hbar \omega - | E_k - E_{k'} |)
     |f_k - f_{k'}|\times \nonumber \\
   & & 
     \left|\int d^3x \Psi_k^{\dag}(x)\nabla_0{\cal H}_0 \Psi_{k'}(x)
         \right|^2 \; .
\eea
  
In the following we discuss the implications of 
$K$, $ F_{\parallel}$, and $ F_{\perp}$ one by one, and show
that Eq.(21) contains both the dissipative effect
and the transverse force, or the Berry phase.

\section{ Discussions}

For the purpose of getting the friction and the transverse force,
the precise value of $K$ in Eq.(22)is irrelevant.

The longitudinal kernel $F_{\parallel}$
contains all information on the vortex friction, revealed by the fact that 
Eqs.(21,23) are identical to  the nonlocal action of Eqs.(5,6).
The friction is determined by the 
low frequency behavior of the spectral function $J(\omega)$, Eq.(28).
  
If we classify the eigenstates of the 
BdG equation, Eq.(17),
according to core(localized) and extended states,
the dissipation comes from all three parts: 
core-to-core, core-to-extended (or extended-to-core), 
and extended-to-extended transition rates.
Those rates are added up, because they are all positive, which
differs from the situation for the transverse kernel.

For a clean superfluid with no impurities, 
it is straightforward to find the damping corresponds to super-Ohmic cases. 
The dissipation is weak in this case.
The main effect is to renormalize the vortex mass.\cite{az}

The presence of impurities mixes up all the clean limit eigenstates, 
and generates a quasi-continuous core energy spectrum
after the impurity average.
A perturbative calculation shows that it is indeed the case,
and the spectral function becomes Ohmic.\cite{az}
This suggests that in the clean limit the dissipation is super-Ohmic, 
and turns into Ohmic in the dirty limit.

In the long time limit 
there are two equivalent forms for the transverse kernel $F_{\perp}$:
The virtual transition expression, Eq.(24), and 
the individual state contribution, Eq.(25). 
Their equivalence has already been discussed in 
Ref.\cite{ao2} with the aid of conservation laws.
Eq.(24) can be again casted into various equivalent forms 
because of the cancelations among those virtual transitions.
For example, at zero temperature only two core states closest to the Fermi
surface, one below `--' and one above `+', will be involved: 
\bea
   F_{\perp} & = & - \partial_{\tau-\tau'} \delta(\tau-\tau') 
                       \int d^3x \int d^3x' \;
         \hbar ( f_{-} - f_{+} ) \nonumber \\
    & &   \hat{z}\cdot
       \left(  \Psi_{-}^{\dag}(x') \nabla_0\Psi_{+}(x')
        \times \nabla_0\Psi_{+}^{\dag}(x)  \Psi_{-}(x) \right) \; .
\eea 
In this case it is a statement of the spectral flow\cite{ao2}.
One may also cast it into the forms of core-to-extended, 
or extended-to-extended transitions.
 
In Eq.(25), the counting of individual state contributions is 
expressed as an area integral of the momentum commutator. 
It can be expressed as the line integral of the momentum density 
far away from the core,
\be
    F_{\perp} =  i \partial_{\tau-\tau'} \delta(\tau-\tau') 
       \oint dx \cdot j(x-x_0 ) \; ,
\ee   
with the momentum density
$j(x) \equiv i \hbar \lim_{x\rightarrow x'} 
 \frac{1}{2} (\nabla -\nabla') \rho_1(x,x')$, 
and the one-body density matrix 
\be
   \rho_1(x,x') = \sum_k \left( ( f_k u_k^{\ast}(x) u_k(x') 
    - (1-f_k) v_k^{\ast}(x) v_k(x') \right)  \; .
\ee
Then $F_{\perp}$ corresponds the term, 
the Berry phase $ i h (\rho_s/2) \int d\tau \; 
\delta \dot{x}_v \cdot (\delta x_v \times \hat{z} )/2$, 
in the effective vortex action Eq.(21). 
Here $\rho_s$ is the superfluid number area density.
The immediate conclusion, as drawn in Ref.\cite{thouless}, is that
localized core states do not contribute, because both 
the total momentum density at the core $x_0$ 
and the momentum density of core states far away from $x_0$ are zero. 
Thus we have two complete different ways to evaluate the transverse force:
via virtual transitions at the core, or a counting of momentum density
for each state far away from the core.
This is similar to the evaluation of the quantum Hall conductance:
the two methods, through either edge or bulk states, are equivalent.  

The insensitive to details such as impurities 
is most transparent from Eq.(25) or (31):
In the one-body density matrix both the electron number 
density and the phase $\theta(x - x_v)$  
are all insensitive to details, if the localization effect caused by
impurities is negligible.
Here the phase $\theta$ is defined through the order parameter 
$\Delta(x,t, x_v) \rightarrow  |\Delta| e^{ i q \theta(x-x_v) } $, with 
$q = \pm 1 $ describing the vorticity along the $\hat{z}$ direction 
and $\theta(x) = \arctan(y/x)$.  

\acknowledgments
{ P.A. is grateful for the hospitality 
 of the Institute for Nuclear Theory in Seattle, where part of 
 the work was done. 
 This work was supported in part by USA DOE(P.A.) and by Swedish NFR. }

\end{document}